# Predicting Performance Under Stressful Conditions Using Galvanic Skin Response

Carter Mundell, Juan Pablo Vielma, and Tauhid Zaman*

Sloan School of Management, Massachusetts Institute of Technology, Cambridge, MA 02139 USA.

*To whom correspondence should be addressed. E-mail: zlisto@mit.edu

**The rapid growth of the availability of wearable biosensors has created the opportunity for using biological signals to measure worker performance. An important question is how to use such signals to not just measure, but actually predict worker performance on a task under stressful and potentially high risk conditions. Here we show that the biological signal known as galvanic skin response (GSR) allows such a prediction. We conduct an experiment where subjects answer arithmetic questions under low and high stress conditions while having their GSR monitored using a wearable biosensor. Using only the GSR measured under low stress conditions, we are able to predict which subjects will perform well under high stress conditions, achieving an area under the curve (AUC) of 0.76. If we try to make similar predictions without using any biometric signals, the AUC barely exceeds 0.50. Our result suggests that performance in high stress conditions can be predicted using signals obtained from wearable biosensors in low stress conditions.**

For many jobs, it is difficult to test a worker's performance under real-life conditions either because it is prohibitively expensive or because the stakes are too high to risk an error. For instance, it can be difficult to simulate the actual pressure and risk faced by a financial trader on a trading floor. The true risk of an accident is hard to recreate for airline pilots or truck drivers. A soldier can only experience the reality of combat on the battlefield. In each of these examples, the worker's performance is evaluated using some sort of testing in a simulation or another low stress environment. For instance, airline pilots and soldiers are often tested in computer-generated simulations (Bymer 2012) (Chang 2013). The performance of a worker in a simulator or in another low stress environment may not be a good predictor of his performance in a high stress situation because it is not obvious how he will respond to the additional stress. Some workers may thrive in high stress environments, while others may suffer a large degradation in their performance. In this paper, show that biometric data, in particular galvanic skin response (GSR), can be used to predict performance under stress and potentially enhance conventional testing methods.

The relationship between stress and performance was first characterized by the famous Yerkes Dodson curve (Yerkes 1908) which posits that performance is maximized at a certain level of stress and decreases as the stress exceeds this optimal level. Other studies have found that this relationship holds for a variety of tasks and types of stress (Ariely 2009, Anderson 1994, S. L. Beilock 2007). Different theories have been proposed to try and explain this relationship. One theory is that added stress causes a shift from automatic to controlled mental processes, resulting in decreased performance (Langer 1979), (Baumeister 1984), (Camerer 2005), (Dandy 2001). Another theory is that the added stress consumes mental resources and interrupts proceduralized routines (S. L. Beilock 2007) . While these theories provide different explanations for the impact of stress on performance, one criticism is that these theories are not predictive because the optimal stress level varies by task and person and also it can be difficult to quantifiably measure stress (Neiss 1988).

One proposed way to measure stress involves using what is known as galvanic skin response (GSR), which is the electrical conductance of the skin and depends on pre-secretory activity in the sweat glands (Lader 1962). GSR was first shown to be correlated with stress in (Jung 1908) and many studies since have supported this relationship (Boucsein 2012, Picard 1997). The recent ubiquity of wearable biosensors such as smart watches and wristbands has made GSR data much easier to collect and has facilitated a new body of experimental research which uses GSR for stress detection. It has been shown to be effective for distinguishing varying degrees of stress in both lab studies (Zhai 2005) and real-life settings, such as highway and city driving (Healey 2005). Experiments have also shown that GSR can be used to infer the difficulty or cognitive load associated with a task (Setz 2010) , (Nourbakhsh 2012).

While the large body of work on GSR has shown that it is correlated with stress and task difficulty, what has not been established is the ability of GSR to predict the performance of a worker under stressful conditions. Being able to predict performance under stress using GSR would make it an incredibly useful signal for a host of applications, such as testing and evaluating workers. Here we show that GSR measured while performing a task under low stress is actually *predictive* of performance of the same task under high stress. That is, by testing a person under a low stress condition, we can predict how well they will perform when there is increased stress.

**Experimental setup**
We conducted a multi-stage experiment with varying degrees of stress, but uniform task difficulty to study the relationship between GSR and performance. We recruited 30 undergraduate students as subjects for the experiment. The subjects were told they would be participating in a study of mathematical ability and stress. Each subject was fitted with a NEULOG GSR finger sensor on the index and middle fingers on their non-dominant hand. The GSR signal was captured throughout the whole experiment at a 5 Hz sampling rate. The subject was first asked to relax for three minutes. After this rest period they began three distinct rounds of a computer-based arithmetic game consisting of a series of three digit by two digit multiplication questions. Between each round the subject was given a three minute rest period. Figure 1 shows a screenshot of the game, the finger sensor used for the experiment and a diagram depicting the sequence of stages in the experiment.

The first round of the game was for calibration and designed to be minimally stressful in order to evaluate the subjects' baseline performance and GSR response to the arithmetic questions. Subjects answered multiplication questions for ten minutes. They did not receive any feedback on whether or not their answers were correct and there was no time limit for answering individual questions. During this round, subjects were incentivized with a small financial reward of $0.25 for each correct response. During the rest period following the calibration round, the subjects' average response time was calculated and used in the following two rounds for the incentive scheme described below.

The second and third rounds of the game were the low and high stress rounds. Both of these rounds consisted of 20 multiplication questions. Subjects were given 130% of their average answer time from the calibration round to answer each question in these rounds. They were instructed that a portion of the time allowed was "bonus" time, during which correct answers were worth $1 instead of $0.25. In the low stress round, 85% of the question time allowed was bonus time, whereas during the high stress round only 50% of the time was bonus time. These percentages were chosen such that achieving the bonus payout in the low stress round was only slightly challenging, whereas in the high stress round it was very difficult. The time remaining for each question was visible to the subjects on the game screen as a bar that reduced in size every second, with bonus and normal time represented by green and red shading respectively, as shown in Figure 1. Subjects were alerted when they ran out of bonus time by an unpleasant buzzer sound. During these two stress rounds, subjects were also given feedback indicating whether the answer they submitted for each question was correct.

### Exploratory Data Analysis and Feature Engineering

Two different performance metrics can be developed from our experiment: accuracy on the 20 questions asked in each round and monetary earnings in each round. In our analysis, we focus on the earnings metric as the dependent variable of interest because earnings is likely to be more sensitive than accuracy to the impact of stress on the subject's performance. In our Supplementary Materials, we include analysis showing that the accuracy is not well predicted by either performance features or GSR features. The earnings depend upon how fast the subject answers the individual questions and how many questions are answered correctly, whereas the accuracy just depends upon the number of questions answered correctly. If a subject is impacted by the stress negatively, this will have a larger impact on their earnings (which is sensitive to speed) than accuracy (which can remain high even if the subject slows down). If we believe that GSR is related somehow to the stress of the subject, and that the stress is impacting performance, then we would expect the good and bad performing subjects to be more easily distinguished in earnings than in accuracy.

The simplest hypothesis is that subjects who perform well in the calibration and low stress rounds will also perform well in the high stress round. If this is the case, then performance features from the calibration and low stress rounds should be highly correlated with earnings in the high stress round. However, we find that these features are not highly correlated with performance in the higher stress rounds. We plot in Figure 2 each subject's performance metrics, both accuracy and earnings, in calibration and low stress rounds versus earnings in the high stress round. There is no clearly visible relationship between the performance metrics across rounds. This is further supported by the correlation coefficient of these metrics, which are not statistically significant at the five percent level for each performance metric (see Supplementary Materials for details). This suggests that to predict performance under high stress conditions, we will likely need more information than simply performance in lower stress conditions. This also supports our speculation that simulation and testing methods designed to measure a candidate's ability in lower stress conditions are not necessarily good indicators of real-world performance by themselves.

We next look at the predictive power of the GSR signal we measured for each subject. Figure 3 shows the trajectory of the raw GSR signal over the course of the experiment for two subjects, who in this case are the highest and lowest earners in the high stress round. In the figure, each round of the trial is separated by a short rest period, demarcated with red lines, and each round and rest period of the experiment is labeled. The GSR signal is measured in microsiemens ($\mu S$) which is a unit of conductance. The figure shows that the GSR signal can exhibit a variety of behaviors, such as upward and downward trends and rapid oscillations. There are also significant differences in the GSR signal path for these two subjects.

We want to determine what features of the GSR signal will predict performance in the high stress round. The GSR signal is very high dimensional, with over 10,000 sampled values per subject. To obtain a useful characterization the GSR signal for each user, we reduced the full high dimensional signal to a set of low dimensional features. One of the feature is *drift* which is defined as the absolute change in the GSR signal from the beginning to the end of a single round in the experiment. Drift measures any slow variations in the GSR signal over the course of a round. Another feature is the *maximum increase* which is the absolute difference in the GSR signal between its minimum point in the round to the end of the round. Maximum increase measures any trend in the GSR signal that is present at the end of the round. We considered other features as well, but none of them proved significant. Details can be found in Supplementary Materials.

### Classifying performance

Our prediction task is to distinguish between subjects who perform well in the high stress round from those who do not using data from the calibration and low stress rounds. This requires us to assign a binary label to each subject indicating if he is a good or bad performer in the high stress round. As a simple labeling scheme, we say a subject is a good performer if his performance metric is strictly above the median value for all subjects, otherwise the subject is a bad performer. We use the median performance metric as a threshold because it is robust to outliers who have extreme values of the metric. We checked the robustness of our findings with respect to the classification threshold and we also found similar significance results for regression models for continuous earnings (see Supplementary Material for details).

To gain a better understanding of our data, we show a scatter plot of the good and bad performers in a two dimensional feature space in Figure 5. We use the features drift and maximum increase because these set of features lead to a good separation of the good and bad performers. In the scatter plot the good and bad performers are indicated with different shaped markers. We see that each type of performer occupies a separate region of this feature space. The bad performers have a smaller maximum increase and a larger drift.

It is not clear why the features we have used in Figure 5 are effective at separating the good and bad performers. From the figure, it can be seen that the bad performers typically have a low maximum increase and a drift that is near zero or negative in the calibration round. This means that during the calibration round the GSR signal does not increase very much. If we follow theories which suggest that increases in GSR indicate an increase in stress or cognitive load, then the bad performers are not suffering very much stress or cognitive load in the calibration round. In contrast, the good performers have a large maximum increase relative to their drift. These subjects are experiencing a higher level of stress or cognitive load. One possible explanation is that the bad performers do not get themselves focused in the absence of any stress, and then when the stress is added, they are not able to perform well. The good performers, on the other hand, are getting themselves mentally focused naturally without any additional stress. Then, when the external stress is added, they are not affected by it because they have already prepared themselves mentally. This is only one possible explanation for what we observe and there may be other explanations. However, we do not focus on them because our goal is not to explain the relationship between GSR and performance, but rather to predict who will perform well under stress. Therefore, we will not discuss these explanations further, but instead build a predictive model for performance under stress.

To evaluate the statistical significance of these apparent relationships, we developed logistic regression models for the above median high stress earnings label shown in Figure 3. For the performance features, we considered accuracy, earnings and average answer time and for the GSR features, we considered drift and maximum increase, all from both calibration and low stress rounds. Table 1 shows estimation results for models with three different feature sets: performance features, GSR features, and performance and GSR features combined. We compared models according to feature significance and Akaike Information Criterion (AIC) and report in the table the two strongest models from each feature set. We see that the models built with performance features alone contain features from the calibration round only—no performance features from the low stress round were found to be significant predictors of high stress earnings. The best performance feature models contain only one weakly significant predictor each. The two best GSR feature models pair drift and maximum increase features from the same period. The calibration period model exhibits both greater statistical significance for both predictors and a lower AIC (which indicates better model fit). Finally, the best combined feature models are the same as the best GSR feature models with the addition of the calibration accuracy feature, which is not statistically significant in either model. However, the AIC for the combined models does improve slightly over the GSR feature models.

### Predictive Power

To estimate how much of an improvement GSR features offer over performance features for classifying good and bad performers, we evaluate the prediction accuracy for the six models shown in Table 1. The predictive power of our models is evaluated using stratified, four-fold cross-validation. This technique involves separating the data into four separate sets or folds with balanced numbers of each outcome class in all folds. The model being tested is fit on three of the four folds and predictions are generated on the fourth held-out fold. This fit and predict process is repeated four times, with a different fold being held out in order to generate a prediction for each data point. Then, we use these holdout predictions to develop receiver operating characteristic (ROC) curve, a visualization of the trade-off between the true positive and false positive rates of your predictions given different values for the decision rule. To assess the predictive accuracy, we calculate the area under the curve (AUC). For context, an AUC of 0.5 indicates that a predictive model is no better than random guessing and the closer an AUC is to 1.0, the more accurate it is considered to be. We generated 500 different permutations of stratified folds to evaluate the results' sensitivity to the data splits and ran the above techniques on each one. Figure 5 shows boxplots for the distribution of AUCs for each model over the 500 permutations. Note from the figure, performance feature models perform no better than random guessing (AUC = 0.5) whereas the GSR feature models achieve a median AUC of 0.76. It is also worth noting that the combined feature models do not appear to achieve higher accuracy than the GSR feature models, suggesting that performance features provide very little if any predictive information not already contained in the GSR features for our data.

## Discussion

Our results demonstrate that GSR signal can be used to predict the performance on cognitive tasks in high stress conditions. An important aspect of our work is the fact that the GSR signal used for prediction is obtained under calibration and low stress conditions. Therefore it is truly predictive of performance under high stress conditions. We do not yet have a clear explanation as to the mechanism of this phenomenon, although we suspect it is related to the fact that good performers will get focused and mentally prepared for a task without the need for additional stress or pressure. Nonetheless, even without a clear explanation for the mechanism, we are still able to accurately predict the good and bad performers using GSR. This suggests that by using wearable biosensors, one can evaluate the performance of workers under dangerous, high risk, high stress conditions using data from safer, low risk, low stress conditions. We anticipate that our experiment will lead to more analysis on the predictive power of GSR for worker performance under stress in a variety of other tasks.

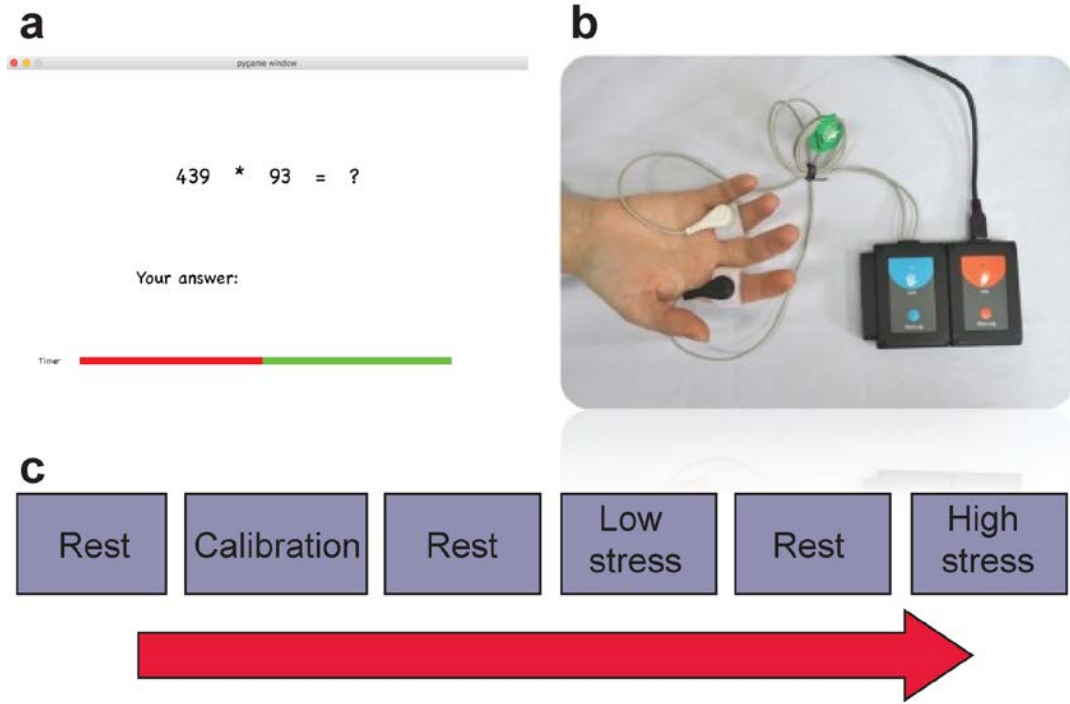

Figure 1. **(a)** A screenshot of the computer interface used for the experiment. **(b)** The Neulog GSR sensor used for the experiment. **(c)** The sequence of stages in the experiment, beginning on the left with a rest stage for the subject and ending on the right with a high stress stage.

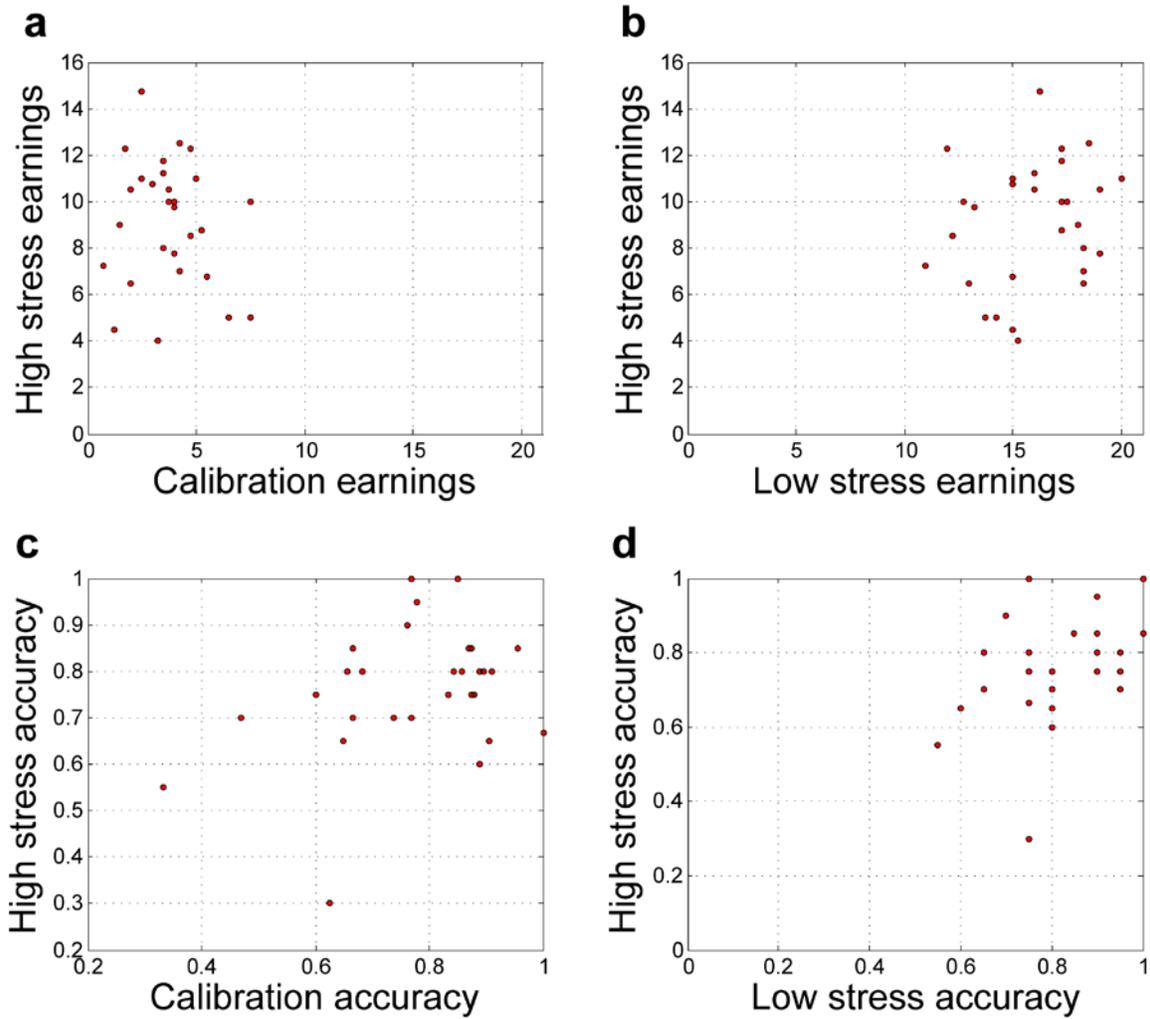

Figure 2. Scatter plots of the performance metrics in different rounds: **(a)** calibration earnings verus high stress earnings, **(b)** low stress earnings versus high stress earnings, **(c)** calibration accuracy versus high stress accuracy, and **(d)** low stress accuracy versus high stress accuracy. The correlation coefficient for each pair of metrics plotted is not significant at the five percent level.

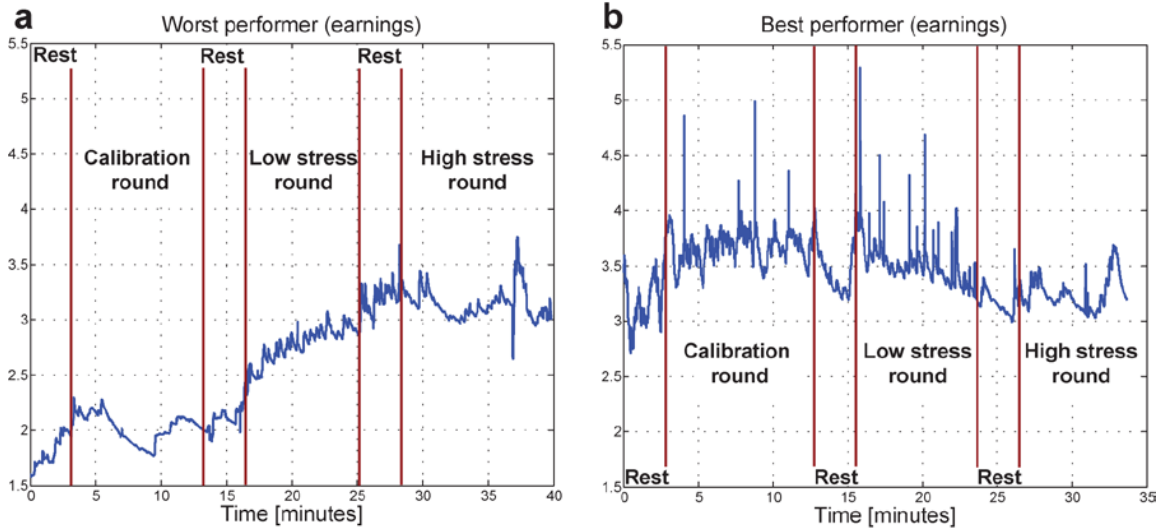

Figure 3. Plots of the GSR signals during the experiment for the highest and lowest earning subjects. The different rounds and rest periods are demarcated by the vertical lines on each plot.

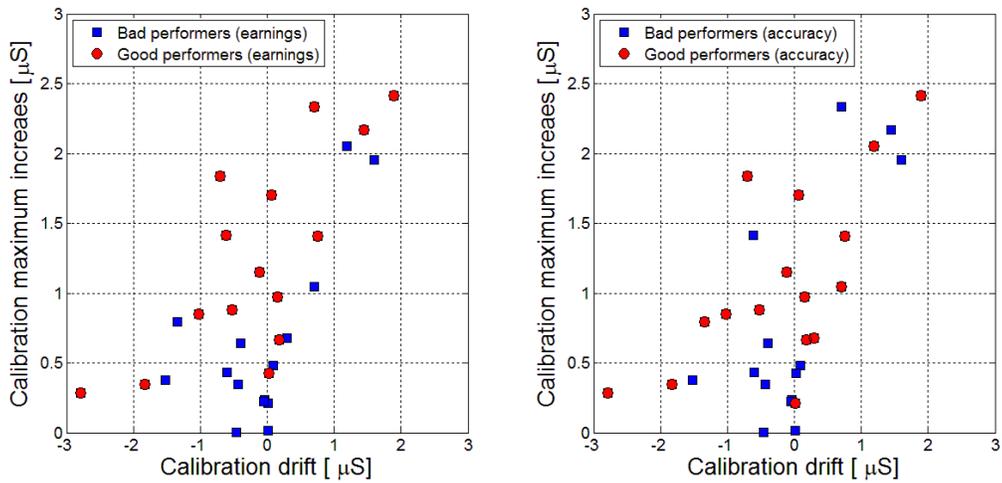

Figure 4. Scatter plots of subjects' calibration drift versus calibration round maximum increase with different markers for good and bad performers in the high stress round in terms of **(a)** earnings and **(b)** accuracy.

| Predictor | Logistic Regression Models of High Stress Performance Dependent Variable: High Stress Earnings ($) | | | | | |
|---|---|---|---|---|---|---|
| | Performance Model #1 | Performance Model #2 | GSR Model #1 | GSR Model #2 | Combined Model #1 | Combined Model #2 |
| Calibration Accuracy | 7.70* (3.86) | 11.08* (4.94) | | | 6.66 (3.95) | 8.46 (4.98) |
| Calibration Answer Time | 0.04 (0.036) | | | | | |
| Calibration Earnings | | -0.65 (0.38) | | | | |
| Calibration GSR Drift | | | -1.39* (0.65) | | -1.47* (0.71) | |
| Calibration GSR Max. Increase | | | 2.82** (1.02) | | 3.01** (1.09) | |
| Low Stress GSR Drift | | | | -1.20 (0.70) | | -1.12 (0.71) |
| Low Stress GSR Max. Increase | | | | 3.27* (1.33) | | 4.01* (1.65) |
| Number of Observations | 30 | 30 | 30 | 30 | 30 | 30 |
| Akaike Information Criterion | 42.17 | 40.04 | 35.56 | 36.92 | 33.62 | 34.33 |

*: p < 0.05, **: p <0.01

Table 1. Predictor coefficients and AIC for six best logistic regression models

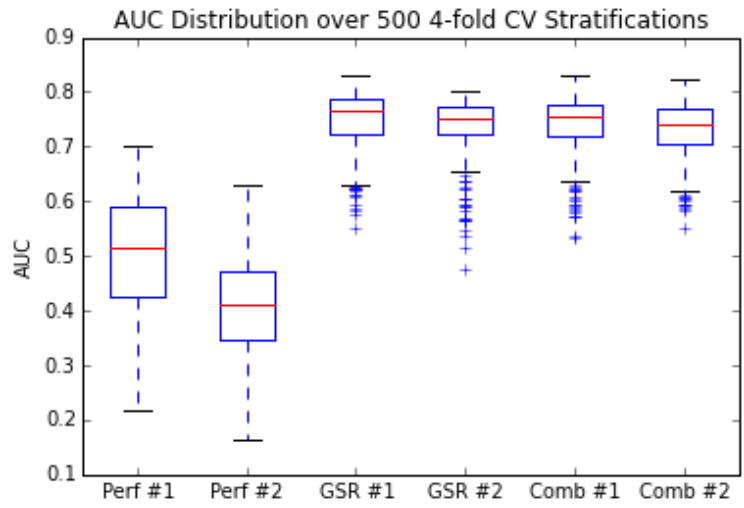

Figure 5. Box plots of models AUCs developed with 500 permutations of stratified, 4-fold cross-validation.